# Optimal Configuration of Wind-to-Ammonia with the Electric Network and Hydrogen Supply Chain: A Case Study of Inner Mongolia

Jiarong Li, *Student Member, IEEE*, Jin Lin, *Member, IEEE*, Philipp-Matthias Heuser, Heidi Ursula Heinrichs, Jinyu Xiao, Feng Liu, *Member, IEEE*, Martin Robinius, Yonghua Song, *Fellow, IEEE*, Detlef Stolten

*Abstract*—Converting wind energy into ammonia (WtA) has been recognized as a promising pathway to enhance the usage of wind generation. This paper proposes a generic optimal configuration model of WtA at the network level to minimize the ammonia production cost by optimizing capacities and locations of WtA facilities including wind turbines, electrolyzers, hydrogen tanks and optimizing supply modes among regions. Specifically, the temporal fluctuation characteristics of wind resources, the operation flexibility of the ammonia synthesis reactor and the transport distances are considered. Three typical supply modes, i.e., the Local WtA, the EN (electric network)-based WtA and the HSC (hydrogen supply chain)-based WtA, combined with two energy transport modes including EN and HT (Hydrogen truck trailers) are included with the consideration of the maximal energy transport capacity of EN and transport distance per day of HT (500km). Real data of Inner Mongolia (a typical province in China with rich wind resources and existing ammonia industries) is employed to verify the effectiveness and significance of proposed model. The effect of above significant factors on optimal planning capacity of WtA facilities and optimal energy transport modes is analyzed, which provides guidelines for WtA configuration. The economic analysis shows that the average LCOA (levelized cost of ammonia) for WtA is approximately 0.57 €/kg in Inner Mongolia and comparable to that for CtA (coal-to-ammonia, 0.41 €/kg) with a reduction of 30% in capacity cost of the facilities.

*Index Terms*—wind-to-ammonia (WtA); wind resources potential evaluation; optimal configuration model

NOMENCLATURE

The main notations in this paper are listed below; other symbols are defined as required.

J. Li is with the State Key Laboratory of Control and Simulation of Power Systems and Generation Equipment, Department of Electrical Engineering, Tsinghua University, Beijing 100084, China and also at the Institute of Techno-Economic Systems Analysis (IEK-3), Forschungszentrum Jülich GmbH, Jülich 52425, Germany (e-mail: jr-li16@mails.tsinghua.edu.cn).

J. Lin and F. Liu are with the State Key Laboratory of Control and Simulation of Power Systems and Generation Equipment, Department of Electrical Engineering, Tsinghua University, Beijing 100084, China.

P.-M. Heuser, H. U. Heinrichs and M. Robinius are with the Institute of Techno-Economic Systems Analysis (IEK-3), Forschungszentrum Jülich GmbH, Jülich 52425, Germany.

D. Stolten is the director of the Institute of Techno-Economic Systems Analysis (IEK-3), Forschungszentrum Jülich GmbH, Jülich 52425, Germany and the Chair for Fuel Cells, RWTH University of Aachen, Germany.

J. Xiao is with the Global Energy Interconnection Development and Cooperation Organization, Beijing 100031, China.

Y. Song is with the State Key Laboratory of Internet of Things for Smart City, University of Macau, Taipa, Macau SAR 999078, China and also with the Department of Electrical Engineering, Tsinghua University, Beijing 100084, China.

A. *Variables*

a. *Operation Variables*

| | |
|---|---|
| $A_i^{\mathrm{L}} / A_i^{\mathrm{E}} / A_i^{\mathrm{H}}$ | Daily ammonia production of **L**ocal WtA/**E**N-based WtA/**H**SC-based WtA in region $i$ |
| $P_{i,t} / E_i$ | Total wind power/daily wind energy in region $i$ |
| $P_{i,t}^{\mathrm{L}} / E_i^{\mathrm{L}}$ | Wind power/Daily wind energy for **L**ocal **Wt**A in region $i$ |
| $P_{i,t}^{\mathrm{ES}} / E_i^{\mathrm{ES}}$ | Wind power/Daily wind energy for **E**N-based WtA in **S**ource region $i$ |
| $P_{i,t}^{\mathrm{HS}} / E_i^{\mathrm{HS}}$ | Wind power/Daily wind energy for **H**SC-based WtA in **S**ource region $i$ |
| $P_{i,t}^{\mathrm{ED}} / E_i^{\mathrm{ED}}$ | Wind power/Daily wind energy for **E**N-based WtA in **D**emand region $i$ |
| $P_{i \to j,t} / E_{i \to j}$ | Wind power/Daily wind energy transported from region $i$ to region $j$ |
| $P_{i,t}^{\mathrm{E}}$ | Wind power injecting to EN from region $i$ |
| $\boldsymbol{P}_{\mathrm{N},t}^{\mathrm{E}} / \boldsymbol{P}_{\mathrm{n},t}^{\mathrm{E}}$ | Vector of node power in EN with/except for the reference node |
| $\boldsymbol{P}_{\mathrm{B},t}$ | Vector of branch power in EN |
| $H_{i \to j}$ | Daily hydrogen transported from region $i$ to region $j$ |
| $\boldsymbol{H}_{\mathrm{B}} / \boldsymbol{H}_{\mathrm{P}}$ | Vector of the daily branch/path hydrogen in the HSC |
| $m_{i,t}^{\mathrm{BUF}}$ | Hydrogen quantity in the buffer tank in region $i$ at time interval $t$ |
| $n_{\mathrm{H}_2,\mathrm{in},i,t}$ $n_{\mathrm{H}_2,\mathrm{out},i,t}$ | Hydrogen input/output flow rate of the buffer tank in region $i$ at time interval $t$ |
| $n_{\mathrm{H}_2,\mathrm{out},i,\min}$ $n_{\mathrm{H}_2,\mathrm{out},i,\max}$ | Minimal/maximal hydrogen output flow rate of the buffer tank in region $i$ |

b. *Planning Variables*

| | |
|---|---|
| $P_i^{\mathrm{RE}}$ | Capacity of the wind turbines in region $i$ |
| $P_i^{\mathrm{EL(L+H)}}$ | Capacity of the electrolyzers for the Local WtA and other regions' HSC-based WtA in region $i$ |
| $P_i^{\mathrm{EL(E)}}$ | Capacity of the electrolyzers for EN-based WtA in region $i$ |
| $m_i^{\mathrm{BUF}}$ | Capacity of the hydrogen buffer tank in region $i$ |
| $m_i^{\mathrm{HS}}$ | Capacity of the hydrogen storage tank in region $i$ |

c. *Economic Variables*

| | |
|---|---|
| $LCOE_i$ | Levelized cost of electricity in region $i$ |
| $LCOH_i$ | Levelized cost of hydrogen in region $i$ |
| $LCOA_i^{\mathrm{L}}$ $LCOA_i^{\mathrm{E}}$ $LCOA_i^{\mathrm{H}}$ | Levelized cost of ammonia for Local WtA/EN-based WtA/HSC-based WtA in region $i$ |
| $EX^f$ | Total cost of facility $f$ per day |
| $CAPEX^f$ | Capacity cost of facility $f$ per day |
| $\text{fix}OPEX^f$ | Fixed operation cost of facility $f$ per day |
| $\text{var}OPEX^f$ | Variable operation cost of facility $f$ per day |
| | $f \in$ {RE (wind turbine), EL (electrolyzer), BUF (hydrogen buffer tank), A (ammonia synthesis reactor), EN, HSC, HT (hydrogen truck trailer), HS (hydrogen storage tank), truck, trailer} |

*B. Indicators, sets and parameters*

| | |
|---|---|
| $s(i)$ | Set of nodes with node $i$ being the start node |
| $e(i)$ | Set of nodes with node $i$ being the end node |
| $i/\mathbb{R}/|\mathbb{R}|$ | Indicator/set/total number of regions |
| $t/\mathbb{T}/|\mathbb{T}|$ | Indicator/set/total number of intraday time intervals |
| $\mathbb{P}$ | Set of paths among the regions |
| $A_i$ | Ammonia production in region $i$ per day |
| $E_i^{\max}$ | Maximal wind energy potential in region $i$ |
| $P_i^{\text{RE,max}}$ | Maximal wind turbine capacity in region $i$ |
| $a_i / b_i$ | Parameters of $E_i$ - $P_i^{\text{RE}}$ in region $i$ |
| $p_{i,t}^{\text{RE}}$ | Typical wind profiles in region $i$ in one day |
| $\eta_{\text{WtA}}$ | Energy conversion coefficient of wind resources to ammonia (17/165 kgNH$_3$/kWh) |
| $\eta_{\text{WtH}}$ | Energy conversion coefficient of wind resources to hydrogen (1/55 kgH$_2$/kWh) |
| $k_{\text{HtA}}$ | Material conversion coefficient of hydrogen to ammonia (17/3 kgNH$_3$/kgH$_2$) |
| $k_{\text{NtA}}$ | Material conversion coefficient of nitrogen to ammonia (17/14 kgNH$_3$/kgN$_2$) |
| $k_{\text{wtH}}$ | Material conversion coefficient of water to hydrogen (1/9 kgH$_2$/kgH$_2$O) |
| $k_{\min} / k_{\max}$ | Coefficient of the minimal/maximal hydrogen output flow rate to ammonia production (0.007/0.01 (kg/hH$_2$)/(kg/dNH$_3$)) |
| $\mathbf{T}_E$ | Node-Branch associate matrix of EN |
| $\mathbf{S}_{PP}$ | Branch-Node sensitivity matrix of EN |
| $\mathbf{T}_H$ | Path-Branch associate matrix of HSC |
| $\mathbf{P}_{B,\min} / \mathbf{P}_{B,\max}$ | Lower/upper limit of the available branch transmission capacity of EN |
| $H_{i \to j,\max}$ | Upper limit of the hydrogen transport capacity in path $i \to j$ of HSC |
| $c_{N_2}$ | Unit cost of nitrogen (0.1 €/kg) |
| $c_{\text{water}}$ | Unit cost of water (0.004 €/kg) |
| $c_{\text{diesel}}$ | Unit cost of gasoline (0.42/4300 €/km/kg) |
| $c_{EN}$ | Wheeling charge of electricity transport via EN (0.008 €/kWh) |
| $D_{j \to i}$ | Distance between region $i$ and region $j$ |

## I. INTRODUCTION

A great quantity of wind resources has not been exploited in the world mainly due to the lack of transmission capacity and consumption. The China Meteorological Administration evaluates that the technical potential of wind power in China is 2,600 GW [1], which corresponds to approximately 7,800 TWh of electricity generation under the assumption of 3,000 full-load hours per year [2], [3]. However, up to the end of 2019, the existing capacity of wind turbines in China is only 210 GW [4]. Meanwhile, the development of traditional hydrogen (H$_2$) -based chemical industry, especially for the ammonia (NH$_3$) industry, which occupies the largest hydrogen downstream market in China, has been facing double pressure from both high fossil fuels consumption and environmental concerns. In 2015, the coal consumption was approximately 80 Mtce (0.67 TWh) to produce ammonia in China, where the CO$_2$ emission was approximately 145 Mt, and the annual growth rate was approximately 4% [5]. A promising approach is to configure the pathway to convert the unexploited wind resources into ammonia (WtA), which provides a kind of considerable consumption to wind resources and simultaneously substitutes fossil fuels in the ammonia industry. WtA has recently attracted worldwide attention [6]. The USA has launched the "Renewable Energy to Fuels through Utilisation of Energy-Dense Liquids" (REFUEL) program to develop scalable technology for WtA, and there have been demonstration projects of renewable energy to ammonia in Japan and Australia.

The WtA process first converts wind resources into electricity via wind turbines and subsequently converts electricity into ammonia (P2A). There are two main routes in the P2A process: in the traditional route, we first convert electricity into hydrogen via electrolyzers and subsequently convert hydrogen into ammonia via Haber-Bosch reaction ( $N_2 + 3H_2 \rightleftharpoons 2NH_3$ ) [7]. The technology of this route is mature in both electrolysis process and ammonia synthesis process. The other route is by the electrochemical ammonia synthesis technology [8], but this technology has not sufficiently matured for industrial application. In this regard, this paper concentrates on the first WtA route.

Existing studies on WtA focus on the energy conversion without considering the energy transport among different regions. [9] and [10] establish a WtA system model consisting of wind turbines, electrolyzers and an ammonia synthesis system and investigate the energy and exergy efficiency and operation robustness of WtA. [11] reports a techno-economic analysis for WtA and analyze the composition of the production cost of green ammonia in several north European countries. The results show different ammonia production costs based on different wind resources. The hydrogen buffer tank is also an important facility before ammonia synthesis reactors in the WtA system to weaken the fluctuation of the flow rate of the gas to avoid catalyst deactivation in the reactors. It is also one of the components of the production cost of ammonia. The capacity of hydrogen buffer tanks actually depends on the operation flexibility of ammonia synthesis reactors; [7] proves that with more operation flexibility of the reactor, the required capacity of the hydrogen buffer tanks decreases, which reduces the total production cost of ammonia. However, none of these studies discussed the capacity of hydrogen buffer tanks in the WtA system considering the actual operation flexibility of reactors based on their operation model.

From the aspects of a province or country, due to the spatial discrepancies between wind resources and ammonia industries, WtA at the network level with the energy transport among different regions is required, which is a gap in the existing WtA studies. There are two kinds of energy supply modes. One mode is in the form of hydrogen via HSC, which first converts wind resources into hydrogen in the source regions, and hydrogen is subsequently transported to the demand regions and converted into ammonia [12]. [13] and [14] studied the operation model of HSC to minimize the hydrogen transport cost. The other mode is in the form of electricity via EN, which first converts wind resources into electricity in the source regions and subsequently transports electricity via an electric network (EN) to the demand regions and converts it into ammonia. The major difference between these two modes relies on different locations of electrolyzers. Our previous study has established the HSC-EN operation model, which considers both energy transport modes to determine the optimal capacities and locations of electrolyzers [15]. Furthermore, in the scope of a province or a country, there are differences in characteristics of wind resources in different regions, which may lead to different production costs of ammonia. However, a gap remains in the existing studies to consider the effects of different characteristics of wind resources on the optimal configuration of WtA.

To fill the gap of energy conversion and transport of WtA at the network level, this paper studies the optimal configuration of WtA considering characteristics of wind resources and spatial discrepancies between wind resources and ammonia industries. The optimal configuration of WtA determines the optimal capacity of wind turbines in source regions and hydrogen buffer tanks in the demand regions and the optimal capacities and locations of electrolyzers, which depend on the energy transport modes. On this basis, the optimal supply modes of WtA are determined. The main contributions are three-fold:

1) A generic optimal configuration model of WtA is proposed to minimize the ammonia production cost at the network level. It simultaneously determines the optimal capacities and locations of facilities and the optimal supply modes among the regions. Three supply modes are counted for: Local WtA, EN-based WtA and HSC-based WtA. The energy transport constraints in capacity and distance of EN and HSC are also considered in this model.

2) Several significant factors are emphasized and considered into the optimal configuration model and analyzed. First, to optimize the capacity configuration of WtA, the characteristics of wind resources are considered based on the evaluation of wind resource potential. On this basis the effect of wind generation fluctuation on the capacity planning of WtA facilities is analyzed, which provides guidelines for the utilization of wind resources in different regions. Then, the operation flexibility of ammonia synthesis reactor is considered based on its operation model in sizing hydrogen buffer tanks. Furthermore, the rule of the optimal energy transport modes and siting of electrolyzers with the transport distance is also obtained.

3) Inner Mongolia, as one of the typical provinces in China with rich wind resources and existing ammonia industry, is selected as the demonstrative example for case studies. The empirical results shows that the average LCOA of WtA is 0.57 €/kg, which is comparable to that of CtA (coal-to-ammonia, 0.41 €/kg) with a 30% reduction in capacity cost of the facilities. The case studies evidently confirm the effectiveness and significance of the proposed methodology, illustrating a great potential from both economic and environmental perspectives.

The remainder of the paper is organized as follows: Section II describes the overall problem from both aspects of unexploited wind resource potential and ammonia industry and the configuration of WtA. Section III formulates the overall optimal configuration model of WtA. In Section IV, case studies based on an industrial system of Inner Mongolia are performed to show the optimal planning results and combination of the three supply modes. The summary and conclusions follow in Section V.

## II. PROBLEM FORMULATION

This section first describes the situation of ammonia industry and unexploited wind resource potential in Inner Mongolia in detail. Then, the generic configuration of WtA, which consists of three WtA supply modes (Local WtA, EN-based WtA and HSC-based WtA) is constructed for such cases with spatial discrepancy.

### A. Situation of Ammonia Industry in Inner Mongolia

Inner Mongolia is one of the major provinces of coal-to-ammonia (CtA) industry in China with an ammonia production of 1.06 Mt in 2018 [16]. The ammonia industry concentrates on Eerduosi (region 6), Alashan (region 12) and Huhehaote (region 1). Fig. 1 shows the spatial distribution of ammonia industry in Inner Mongolia.

### B. Situations of Wind Resources in Inner Mongolia

Inner Mongolia is a province in the north of China with an area of 1.18 million km$^2$. It is one of the wind-richest provinces in China with an average wind resource density above 200 W/m$^2$, and the existing capacity of wind turbines is 30.07 GW (2019) [17]. However, a great quantity of wind resources remains unexploited according to the following evaluations.

*a. Evaluation Method of Wind Resource Potential*

In this paper, we evaluate the wind resources potential of each region in Inner Mongolia based on the method proposed by Ryberg et al. [18], [19]. The evaluation results include the locations and time series output of the satisfied wind turbines, based on which the total wind turbine capacity and wind energy potential in each region can be obtained.

*b. Evaluation Results of Inner Mongolia*

Considering the utilization rate of facilities and total energy required by the ammonia production, the threshold of minimum FLH (full-load hour) is set as 4000 h, and there is a total wind capacity potential of approximately 60 GW with ~246 TWh of annual wind energy in Inner Mongolia (all evaluation results are shown in Fig. 9 in the Appendix). Here, we suppose that the existing wind turbines occupy the best wind resources in each region. The spatial distribution of maximal unexploited wind turbine capacity potential in Inner Mongolia is shown in Fig. 2. Regions 2 and 11 have the richest unexploited wind resources.

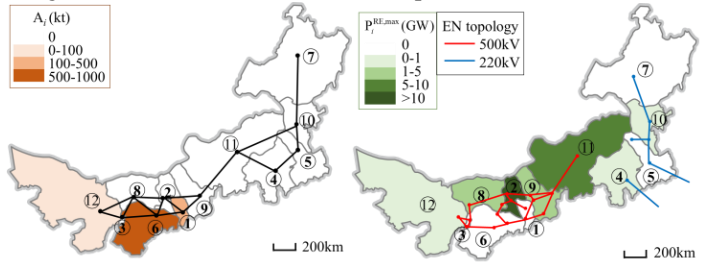

Fig. 1 Distribution of annual ammonia production in Inner Mongolia (kt) and HSC topology

Fig. 2 Distribution of maximal unexploited wind capacity in Inner Mongolia (GW) and EN topology

*c. Characteristics of wind resources*

The differences in characteristics of wind resources mainly involve two aspects: the different FLH, which leads to different wind energy production with the same capacity of wind turbines, and the intraday fluctuation of wind power.

Based on the evaluation results of wind resources in each region, the relationship of $E_i$ with $P_i^{RE}$ in each region can be described by fitting with the quadratic function. The slope of secants of points on $E_i$ - $P_i^{RE}$ relationship represents the average FLH of wind turbines in one day, which gradually decreases with $P_i^{RE}$ increasing. Planning capacity $P_i^{RE}$ should not exceed the maximal unexploited wind turbine capacity potential $P_i^{RE,max}$. (1) and (2) describe the planning model of wind turbine capacity, which belongs to the planning model of WtA in Section III.

$$E_i \leq a_i P_i^{RE^2} + b_i P_i^{RE} \quad \forall i \in \mathbb{R} \tag{1}$$

$$0 \leq P_i^{RE} \leq P_i^{RE,max} \quad \forall i \in \mathbb{R} \tag{2}$$

To be consistent with the models in Section III, here, we use $p_{i,t}^{RE}$, which represents the hourly wind power per unit wind energy generation, to describe the typical wind profiles. Thus, the intraday wind power $P_{i,t}$ can be described by (3) and used in (5).

$$P_{i,t} = E_i p_{i,t}^{RE} \quad \forall i \in \mathbb{R}, t \in \mathbb{T} \tag{3}$$

Fig. 3 is the violin chart that describes the data distribution, their probability density, lower quartile, median, and upper quartile of the strongest intraday wind power fluctuation profiles in the entire year in each region according to the simulation results. The narrower and longer filled area represents the stronger intraday fluctuation. The fluctuation is strong in regions 2, 11, and 12 and slight in regions 4 and 8. The intraday fluctuation may affect the capacity of facilities such as electrolyzers and hydrogen buffer tanks, which is discussed in detail in the case studies.

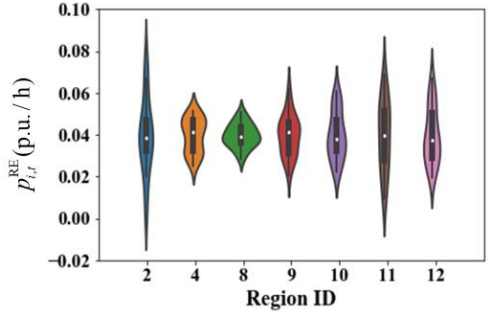

Fig. 3 Violin chart of typical wind profiles in Inner Mongolia

### C. Configuration of WtA

As shown in Fig. 1 and Fig. 2, in cases of spatial discrepancies between the wind resource potential (regions 2, 11, 9) and the ammonia industries (region1, 6, 12) in Inner Mongolia, a generic configuration of WtA that consists of three supply modes (Local WtA, EN-based WtA and HSC-based WtA) is constructed, and two types of energy transport modes are considered: EN and HT (hydrogen truck trailers). Here, we do not consider the ammonia transport mode due to the regulation on ammonia industries, which limits that ammonia production is highly centralized and can only be produced and consumed in regions 1, 6, and 12. For *Local WtA*, a basic *LCOA* composition includes the cost from wind turbines, electrolyzers, hydrogen buffer tanks and ammonia synthesis reactors. For *EN-based WtA*, a fixed wheeling charge of unit electricity transport via EN is considered, which comes from the EN operators. For *HSC-based WtA*, the extra cost via HSC comes from the investment, operation and maintenance costs for trucks, trailers and hydrogen storage tanks. For HSC-based WtA, the cost of the hydrogen buffer tanks is not considered, since hydrogen storage tanks for daily hydrogen storage are considered. The detailed calculation methods of the three described *LCOA*s are introduced in Section III.

The distribution of ammonia industries and wind resources and the topology of HSC and EN in Inner Mongolia are shown in Fig. 1 and Fig. 2. According to this distribution, there should only be Local WtA in region 12, since there is no wind resource potential in regions 1 and 6. Ammonia production in regions 1 and 6 may be satisfied by EN-based WtA from regions 2, 8, 9, 11 and 12 or HSC-based WtA from all other source regions. The optimal combination of three supply modes mainly depends on the characteristics of wind resources and the energy transport costs and constraints of EN and HSC. All of these considerations are included in the optimal configuration model of WtA in Section III.

## III. OPTIMAL CONFIGURATION MODEL OF WTA

This section introduces the optimal configuration model of WtA, including the generic operation model, planning model and calculation methods of three *LCOA*s to minimize the total ammonia production cost.

### A. Generic Operation Model of WtA

To better illustrate the model, Fig. 4 shows a simple system of two regions: source region *i*, where wind energy $E_i$ is higher than the energy required in ammonia production $A_i$, and demand region *j*, where wind energy $E_j$ is lower than the energy required in ammonia production $A_j$.

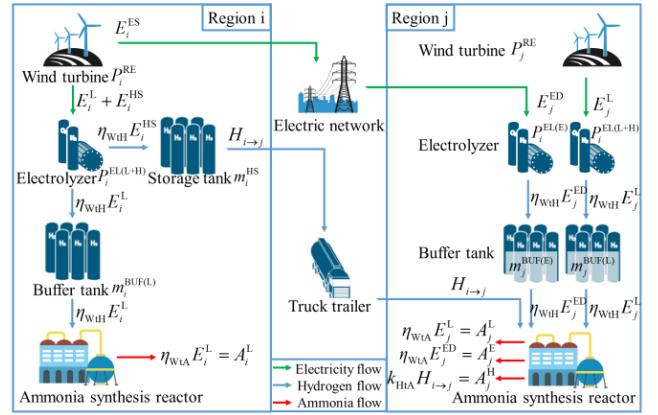

Fig. 4 Illustration of the configuration of WtA and three supply modes

The time increment of the ammonia production and operation of the HSC is one day [20], while the intraday operation of EN is resolved hourly ($\tau$=1 h). Based on Fig. 4, for an arbitrary region *i*, the combination of wind energy and ammonia production based on the three defined WtA supply modes can be expressed as (4) and (5).

$$A_i = A_i^L + A_i^E + A_i^H \quad \forall i \in \mathbb{R} \tag{4}$$

$$P_{i,t} = P_{i,t}^L + P_{i,t}^{ES} + P_{i,t}^{HS} \quad \forall i \in \mathbb{R}, t \in \mathbb{T} \tag{5}$$

For an easier expression, we define the operation parameters of wind energy with the time interval of one day, and the relationship of $P_{i,t}$ and $E_i$ is introduced in (3).

$$E_i^L = \sum_{t=1}^{|\mathbb{T}|} P_{i,t}^L \tau, E_i^{ED} = \sum_{t=1}^{|\mathbb{T}|} P_{i,t}^{ED} \tau, E_i^{HS} = \sum_{t=1}^{|\mathbb{T}|} P_{i,t}^{HS} \tau \quad \forall i \in \mathbb{R}$$

$$E_{i \to j} = \sum_{t=1}^{|\mathbb{T}|} P_{i \to j,t} \tau \quad \forall i \to j \in \mathbb{P} \tag{6}$$

The relationship of ammonia production and wind energy of three WtA supply modes in region *i* satisfies:

$$A_i^L = \eta_{WtA} E_i^L \quad \forall i \in \mathbb{R} \tag{7}$$

$$A_i^E = \eta_{WtA} E_i^{ED} \quad \forall i \in \mathbb{R} \tag{8}$$

$$A_i^H = \sum_{j \in e(i)} k_{HtA} H_{j \to i} \quad \forall i \in \mathbb{R} \tag{9}$$

$$\eta_{\text{WtH}} E_i^{\text{HS}} = \sum_{k \in s(i)} H_{i \to k} \quad \forall i \in \mathbb{R} \tag{10}$$

The wind power of region $i$ for EN-based WtA is the sum of the power transported to/from other regions:

$$P_{i,t}^{\text{ES}} = \sum_{j \in s(i)} P_{i \to j,t}, P_{i,t}^{\text{ED}} = \sum_{j \in e(i)} P_{j \to i,t} \quad \forall i \in \mathbb{R}, t \in \mathbb{T} \tag{11}$$

The energy transport via EN and HSC satisfy the operation model of EN and HSC. In our previous research [15], there is the detailed HSC-EN operation model, which is directly used in this paper. Here, we only list the important equations as follows.

*a. EN*

Here, we consider the existing infrastructure of EN. The KCL constraint of each node should be satisfied, which can be described with node-branch associate matrix $\mathbf{T}_E$. Based on the classical DC flow model, which is normally utilized in research on transmission network [21], the relationship of the power increment of transmission lines with the injecting power of the nodes can be described by sensitivity matrix $\mathbf{S}_{PP}$.

$$P_{i,t}^{E} = P_{i,t}^{\text{ES}} - P_{i,t}^{\text{ED}} \quad \forall i \in \mathbb{R}, t \in \mathbb{T} \tag{12}$$

$$\mathbf{P}_{N,t}^{E} - \mathbf{T}_E \Delta \mathbf{P}_{B,t} = 0 \quad \forall t \in \mathbb{T} \tag{13}$$

$$\Delta \mathbf{P}_{B,t} - \mathbf{S}_{PP} \mathbf{P}_{n,t}^{E} = 0 \quad \forall t \in \mathbb{T} \tag{14}$$

The branch power increment should not exceed the available transmission capacity determined by the maximal transmission capacity of the transmission lines and existing power flow.

$$\mathbf{P}_{B,\min} \leq \Delta \mathbf{P}_{B,t} \leq \mathbf{P}_{B,\max} \quad \forall t \in \mathbb{T} \tag{15}$$

*b. HSC*

Here, we model the hydrogen transportation network based on Moore neighborhood model, which is generally used to model the HSC [22], [23]. (16) describes the relationship of paths between any two regions and branches between any adjacent regions based on the shortest route set.

$$\mathbf{H}_B = \mathbf{T}_H \mathbf{H}_P \tag{16}$$

Considering the maximal traveling distance per day of the truck, hydrogen transportation is not considered on path $i \to j$, whose shortest distance is larger than the maximal daily traveling distance.

$$H_{i \to j} \leq H_{i \to j,\max} \quad \forall i \to j \in \mathbb{P} \tag{17}$$

*B. Planning Model of WtA*

The configuration of WtA requires new planning for related facilities including wind turbines, electrolyzers, hydrogen buffer tanks and hydrogen storage tanks. The planning model of wind turbines is introduced in (1)-(2).

*a. Electrolyzer*

For Local WtA and HSC-based WtA, the total capacity of electrolyzers $P_i^{\text{EL(L+H)}}$ is determined by the maximum value $P_{i,t}^{L} + P_{i,t}^{\text{HS}}$ of all time intervals in one day considering the fluctuation of wind energy:

$$P_i^{\text{EL(L+H)}} = \max\{P_{i,t}^{L} + P_{i,t}^{\text{HS}}\} \quad \forall i \in \mathbb{R}, t \in \mathbb{T} \tag{18}$$

For EN-based WtA, the capacity of electrolyzers $P_i^{\text{EL(E)}}$ depends on the maximum $P_{i,t}^{\text{ED}}$ of all time intervals in one day:

$$P_i^{\text{EL(E)}} = \max\{P_{i,t}^{\text{ED}}\} \quad \forall i \in \mathbb{R}, t \in \mathbb{T} \tag{19}$$

*b. Hydrogen Buffer Tank*

For Local WtA and EN-based WtA, hydrogen buffer tanks are required to weaken the fluctuation of the hydrogen flow rate to the acceptable range of the following ammonia synthesis reactor. In this paper, we consider the operation flexibility of the reactor to optimize the capacity of buffer tanks. Based on our previous research [24], we obtain the acceptable range ($n_{H_2,\text{out},i,\min}, n_{H_2,\text{out},i,\max}$) of the output hydrogen flow rate of the buffer tank and the input of the ammonia synthesis reactor, which are approximately proportional to the ammonia production in region $i$. Detailed coefficients $k_{\min}$ and $k_{\max}$ are determined by the parameters of the reactor [24]. The intraday operation model of the buffer tank determines capacity $m_i^{\text{BUF(L)}}$ and $m_i^{\text{BUF(E)}}$ as follows. Here, (20a), (20c), (20f) and (20g) are generic equations for both Local WtA and EN-based WtA.

$$m_{i,t}^{\text{BUF}} = m_{i,t-1}^{\text{BUF}} + n_{H_2,\text{in},i,t-1} - n_{H_2,\text{out},i,t-1} \quad \forall i \in \mathbb{R}, t \in \mathbb{T} \tag{20a}$$

$$n_{H_2,\text{in},i,t} = \begin{cases} \eta_{\text{WtH}} P_{i,t}^{L} \tau & \text{(Local WtA)} \\ \eta_{\text{WtH}} P_{i,t}^{\text{ED}} \tau & \text{(EN-based WtA)} \end{cases} \quad \forall i \in \mathbb{R}, t \in \mathbb{T} \tag{20b}$$

$$n_{H_2,\text{out},i,\min} \leq n_{H_2,\text{out},i,t} \leq n_{H_2,\text{out},i,\max} \quad \forall i \in \mathbb{R}, t \in \mathbb{T} \tag{20c}$$

$$n_{H_2,\text{out},i,\min} = \begin{cases} k_{\min} A_i^{L} & \text{(Local WtA)} \\ k_{\min} A_i^{E} & \text{(EN-based WtA)} \end{cases} \quad \forall i \in \mathbb{R}$$

$$n_{H_2,\text{out},i,\max} = \begin{cases} k_{\max} A_i^{L} & \text{(Local WtA)} \\ k_{\max} A_i^{E} & \text{(EN-based WtA)} \end{cases} \quad \forall i \in \mathbb{R} \tag{20d}$$

$$m_{i,0}^{\text{BUF}} = m_{i,|\mathbb{T}|}^{\text{BUF}} \quad \forall i \in \mathbb{R} \tag{20f}$$

$$m_{i,t}^{\text{BUF}} \leq m_i^{\text{BUF}} \quad \forall i \in \mathbb{R}, t \in \mathbb{T} \tag{20g}$$

*c. Hydrogen Storage Tank*

For HSC-based WtA, hydrogen storage tanks are required to temporarily store hydrogen in region $i$. This part of hydrogen will be later transported to other regions. Thus, the capacity of the storage tanks are determined by the hydrogen quantity transported from region $i$ to other regions per day.

$$m_i^{\text{HS}} = \sum_{j \in s(i)} H_{i \to j} \quad \forall i \in \mathbb{R} \tag{21}$$

*C. Calculation Methods of Three LCOAs*

$LCOE_i$ of region $i$ is defined as the levelized electricity generation cost in (22), which is related to the cost from wind turbines (1000 €/kW [14]). The general calculation method of capital cost *CAPEX* and fixed operation cost fix*OPEX* in (22)-(26) for facilities is obtained from Heuser et al. [25].

$$LCOE_i = \frac{EX_i^{\text{RE}}}{E_i} = \frac{CAPEX_i^{\text{RE}} + \text{fix}OPEX_i^{\text{RE}}}{E_i} \quad \forall i \in \mathbb{R} \tag{22}$$

$LCOH_i$ of region $i$ is defined as the hydrogen production cost based on the local wind energy ($E_i^{L} + E_i^{\text{HS}}$) and determined by the capital and operation cost of $P_i^{\text{EL(L+H)}}$ (500 €/kW [14]):

$$LCOH_i = LCOE_i / \eta_{\text{WtH}} + \frac{EX_i^{\text{EL(L+H)}}}{\eta_{\text{WtH}}(E_i^{L} + E_i^{\text{HS}})} \tag{23a}$$

$$EX_i^{\text{EL(L+H)}} = CAPEX_i^{\text{EL(L+H)}} + \text{fix}OPEX_i^{\text{EL(L+H)}} + \text{var}OPEX_i^{\text{EL(L+H)}} \tag{23b}$$

$$\text{var}OPEX_i^{\text{EL(L+H)}} = \eta_{\text{WtH}}(E_i^{L} + E_i^{\text{HS}})c_{\text{water}} / k_{\text{wtH}} \quad \forall i \in \mathbb{R} \tag{23c}$$

Here, we mainly consider the water cost [14] as the variable operation cost of the electrolyzer except for the electricity cost. Based on the definitions of $LCOE_i$ and $LCOH_i$, three $LCOA$s are defined with the additional costs of buffer tanks, ammonia reactors, EN or HSC as follows.

*a. Local WtA*

The detailed expression of $LCOA_i^L$ is (24). Here, we consider the cost of nitrogen during the ammonia synthesis process as the variable operation cost, and the unit cost of hydrogen tanks is 500 €/kg H$_2$ [26].

$$LCOA_i^L = LCOH_i / k_{HtA} + \frac{EX_i^{BUF(L)} + EX_i^{A(L)}}{A_i^L} \quad (24a)$$

$$EX_i^{BUF(L)} = CAPEX_i^{BUF(L)} + \text{fix}OPEX_i^{BUF(L)} \quad (24b)$$

$$EX_i^{A(L)} = OPEX_i^{A(L)} = c_{N_2} A_i^L / k_{NtA} \quad \forall i \in \mathbb{R} \quad (24c)$$

*b. EN-based WtA*

The detailed expression of $LCOA_i^E$ is (25). The calculation methods of $EX_i^{EL(E)}$, $EX_i^{BUF(E)}$ and $EX_i^{A(E)}$ are identical to $EX_i^{EL(L+H)}$, $EX_i^{BUF(L)}$ and $EX_i^{A(L)}$ in (23) and (24).

$$LCOA_i^E = \frac{EX_i^{RE(E)} + EX_i^{EL(E)} + EX_i^{EN} + EX_i^{BUF(E)} + EX_i^{A(E)}}{A_i^E} \quad (25a)$$

$$EX_i^{RE(E)} = \sum_{j \in e(i)} LCOE_j E_{j \to i} \quad (25b)$$

$$EX_i^{EN} = \sum_{j \in e(i)} c_{EN} E_{j \to i} \quad \forall i \in \mathbb{R} \quad (25c)$$

*c. HSC-based WtA*

The detailed expression of $LCOA_i^H$ is (26). The calculation method of $EX_i^{A(H)}$ is identical to $EX_i^{A(L)}$ in (24). The extra cost of $EX_i^{HSC}$ includes hydrogen transport cost $EX_i^{HT}$ (37.21 €/kg H$_2$ for truck and 200 €/kg H$_2$ for trailer [26]), storage cost $EX_i^{HS}$ (500 €/kg H$_2$ [26]), and variable operation cost of $EX_i^{HT}$ mainly from diesel consumed by trucks in (26f)

$$LCOA_i^H = \frac{EX_i^{RE(H)} + EX_i^{EL(H)} + EX_i^{HSC} + EX_i^{A(H)}}{A_i^H} \quad (26a)$$

$$EX_i^{RE(H)} + EX_i^{EL(H)} = \sum_{j \in e(i)} \left[ LCOH_j H_{j \to i} \right] \quad (26b)$$

$$EX_i^{HSC} = EX_i^{HT} + EX_i^{HS} \quad (26c)$$

$$EX_i^{HT} = \sum_{j \in e(i)} \left[ EX_{j \to i}^{truck} + EX_{j \to i}^{trailer} \right] \quad (26d)$$

$$EX_{j \to i}^{truck} = CAPEX_{j \to i}^{truck} + \text{fix}OPEX_{j \to i}^{truck} + \text{var}OPEX_{j \to i}^{truck} \quad (26e)$$

$$\text{var}OPEX_{j \to i}^{truck} = c_{diesel} D_{j \to i} H_{j \to i} \quad (26f)$$

$$EX_{j \to i}^{trailer} = CAPEX_{j \to i}^{trailer} + \text{fix}OPEX_{j \to i}^{trailer} \quad (26g)$$

$$EX_i^{HS} = CAPEX_i^{HS} + \text{fix}OPEX_i^{HS} \quad \forall i \in \mathbb{R} \quad (26h)$$

*D. Optimal Configuration Model of WtA*

The objective function is to minimize the total ammonia production cost by optimizing the combination of three WtA supply modes in each region. The whole optimization model is as follows:

$$\min \sum_{i=1}^{|\mathbb{R}|} LCOA_i^L \cdot A_i^L + LCOA_i^E \cdot A_i^E + LCOA_i^H \cdot A_i^H \quad (27)$$

$$s.t. \quad (1) - (26)$$

Constraints (1)-(26) comprise the operation model, planning model and economic expressions, which are introduced as above. It is an NLP optimization problem, which can be solved by FICO Xpress Optimization Solver 8.8.1 with the successive linear programming algorithm [27].

## IV. CASE STUDIES

In this section, case studies are performed based on the industrial system of Inner Mongolia. The effect of different wind resources on planning results and the effect of different distances on the energy transport results are first discussed. Then, the optimal configurations of three WtA supply modes and three *LCOA*s are analyzed. Finally, WtA is compared with traditional CtA from the aspects of economic and environmental indices.

*A. Case Descriptions*

The distributions of ammonia industries and unexploited wind resource potential in Inner Mongolia are shown in Fig. 1 and Fig. 2. Except for regions 1, 3, 5, 6, and 7, all other regions have available wind resources with FLH>4000 h, and the ammonia production concentrates in regions 1, 6 and 12. The topology of HSC and EN is also shown in Fig. 1 and Fig. 2. According to existing EN operators in Inner Mongolia, the western 8 regions and the eastern 4 regions belong to different operators, and there is no electricity exchange between them. Detailed values of the economic indices are shown in the Nomenclature, which are obtained from [14], [15], and [24]. The parameters in (1), (2), and transport distances $D_{j \to i}$ in (26f) and other economic indices are listed in Table III, IV and V in the Appendix.

*B. Optimal Planning Results with Different Wind Resources*

*a. Optimal Planning Results of WtA Facilities*

Table I shows the optimal planning results of facilities. There is no planning facility in the regions that are not mentioned. For $P^{RE}$, regions 2, 8, 9, 11 and 12 have the planning wind turbines of 2545 MW in total. The reason of the lack of wind turbine in other regions is explained in subsection *C*. Region 8 has the maximal wind turbine capacity, which occupies 94% of the total planning capacity. For $P^{EL}$, the total capacity is 1574 MW, where 99.98% (1573.87 MW) is located in the source regions, and only 0.02% (0.35 MW) is located in the demand regions. For $m^{BUF}$, 0.01 t buffer tanks in region 1 are required for EN-based WtA with 0.09 t daily hydrogen production, and 0.46 t buffer tanks in region 12 are required for Local WtA with 8.70 t daily hydrogen production. A capacity with no more than 10% of daily hydrogen production is required, since the operation flexibility of the ammonia synthesis reactor is considered. Finally, for $m^{HS}$, 512.67 t storage tanks are required in source regions 2, 8, and 9 for the daily hydrogen transport to demand regions. In summary, there are 2545 MW of wind turbines, 1573.87 MW of electrolyzers and 512.67 t of hydrogen storage tanks in the source regions, and 0.35 MW of electrolyzers and 0.47 t of hydrogen buffer tanks in the demand regions to satisfy approximately 3000 t/d of ammonia production in Inner Mongolia.

TABLE I
OPTIMAL PLANNING RESULTS OF FACILITIES

| Region | $P^{RE}$ (MW) | $P^{EL(L+H)}$ (MW) | $P^{EL(E)}$ (MW) | $m_{BUF}$ (t) | $m_{HS}$ (t) |
|---|---|---|---|---|---|
| 1 | 0 | 0 | 0.35 | 0.01 | 0 |
| 2 | 5 | 0.71 | 0 | 0 | 0.16 |
| 8 | 2385 | 1450.56 | 0 | 0 | 488.41 |
| 9 | 110 | 88.80 | 0 | 0 | 24.10 |
| 11 | 5 | 0 | 0 | 0 | 0 |
| 12 | 40 | 33.80 | 0 | 0.46 | 0 |

*b. Influence of Different Wind Resources on the Planning Results*

Furthermore, comparing planning $P^{RE}$ in Table I with maximal wind turbines' potential $P^{RE, \max}$, we find that although the wind resource is the richest in region 2, the capacity of the planning wind turbines is small. To explain the reason for it, according to the proposed planning model of WtA, we can calculate the required capacity of electrolyzers and hydrogen buffer tanks for the utilization of wind energy in source regions 2, 8, 9, and 11 and the related *LCOA* for Local WtA. The results are shown in Fig. 5. The filled quadrilateral in the main figure consists of four points with $E$=30000 MWh/d in four regions, and in the sub figure, it represents the surface with *LCOA*=0.5124 €/kg. Due to the strong fluctuation of wind energy in region 2, the required capacity of $P^{EL}$ and $m^{BUF}$ is much higher than that in region 8 with the utilization of the same wind energy, which leads to higher *LCOA*. In other words, to attain the same *LCOA* in different regions, a stronger fluctuation corresponds to smaller utilization of wind energy and capacity of the planning wind turbines. Table II shows the final utilization of wind energy $E$ and required capacity of facilities in source regions 2, 8, 9, and 11. The differences in *LCOE* are obvious due to the different utilization of wind energy. However, *LCOA*$^L$s are similar to the addition of costs from $P^{EL}$ and $m^{BUF}$. The small differences in *LCOA*$^L$ are mainly due to different $EX^{EN}$ or $EX^{HSC}$ from these source regions to demand regions.

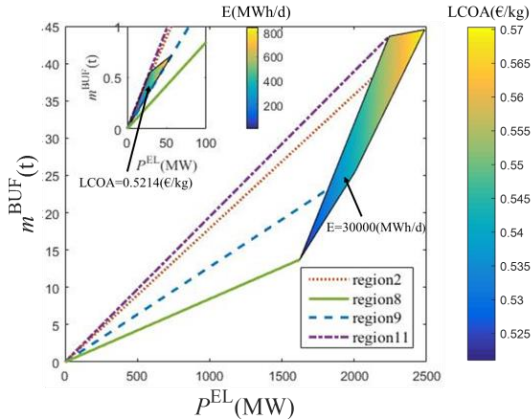

Fig. 5 Relationship of $P^{EL}$ and $m^{BUF}$ with $E$ and the related *LCOA*$^L$

TABLE II
PLANNING RESULTS IN SOURCE REGIONS

| Region | $E$ (MWh/d) | $P^{RE}/P^{RE,max}$ (%) | $P^{EL}$ (MW) | *LCOE* (€/kWh) | *LCOA*$^L$ (€/kg) |
|---|---|---|---|---|---|
| 2 | 9 | 0.02 | 0.71 | 0.0279 | 0.5224 |
| 8 | 26862 | 89.83 | 1450.56 | 0.0337 | 0.5199 |
| 9 | 1325 | 4.91 | 88.80 | 0.0313 | 0.5229 |
| 11 | 5 | 0.07 | 0.35 | 0.0258 | 0.4876 |

### C. Optimal Energy Transport Modes with Different Distances

Except for region 12 where there is local wind resources, the energy required for ammonia production in regions 1 and 6 can only be supplied by other source regions such as regions 2, 8, 9, and 11 via EN or HSC. Fig. 6 shows the optimal electricity and hydrogen supply results between source regions and demand regions. Based on the assumption of constraints in HT with the maximal transport distance being 500 km (50 km/h*10 h [26]) and existing EN operators, the wind energy in the eastern 4 regions (4, 5, 7, and 10) cannot be utilized, so there is no planning wind turbine in these regions (the diameter of the blue circle in Fig. 6 is 500 km). The energy required in region 1 is supplied by region 11 via EN and regions 2, 8, and 9 via HSC. The energy required in region 6 is satisfied by region 8 via HSC. Region 8 also supplies energy to region 12 via HSC, since the local wind energy is not sufficient for its ammonia production. Since $EX^{EN}$ is a fixed value determined by EN operators, while $EX^{HSC}$ is related to supply distances, Fig. 6 shows that the energy required in the demand regions is always supplied by the nearest source regions to reduce the energy transport cost. Thus, $EX^{HSC}$ is lower than $EX^{EN}$ in short distances (below 500 km); for energy supply above 500 km, the electricity supply mode must be the supplement mode since there is constraint on transport distance in HSC.

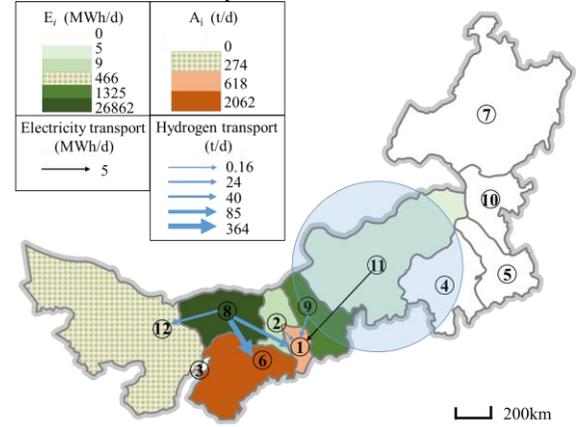

Fig. 6 Optimal energy transport results

### D. Economic Analysis of Three WtA Supply Modes

Fig. 7 shows that there are "EN-based WtA" and "HSC-based WtA" in region 1, which provide 0.08% and 99.92% energy required in ammonia production, respectively. In region 6, there is only "HSC-based WtA". In region 12, there are "Local WtA" and "HSC-based WtA" which provide 18% and 82% energy required in ammonia production, respectively. Fig. 7 shows the compositions of three *LCOA*s in regions 1, 6, and 12. In general, *LCOA* for mode "Local WtA" is obviously lower than *LCOA* for modes "EN-based WtA" and "HSC-based WtA" without the extra cost $EX^{EN}$ or $EX^{HSC}$. The local *LCOA* in region 12 is 0.55 €/kg, $EX^{EN}$ is 0.08 €/kg, and $EX^{HSC}$ is 0.05 €/kg on average. Although there are differences in detailed compositions in *LCOA*s for different supply modes and different regions, the final *LCOA*s are approximately identical (0.57 €/kg) with the optimization of the proposed configuration model considering the differences in wind resources in different source regions and energy transport cost via EN and HSC.

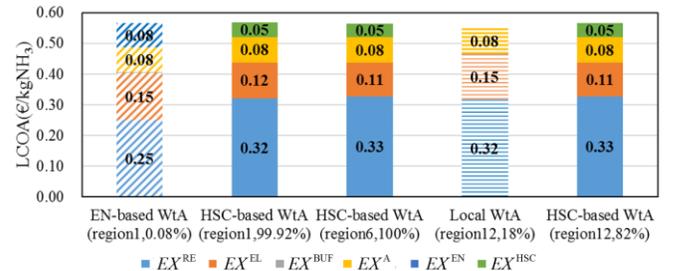

Fig. 7 Composition of *LCOA* in each region in Inner Mongolia

### E. Comparison with CtA

The average *LCOA* for WtA in Inner Mongolia is 0.57 €/kg, while the average *LCOA* for CtA considering the carbon tax of approximately 25 €/t [28] is only 0.41 €/kg [29]. According to Fig. 7, this is mainly due to the relatively expensive capacity cost of wind turbines (1000 €/kW) and electrolyzers (500 €/kW), which contribute 58% and 27% of *LCOA* in the mode "Local WtA" in region 12 for example. We analyze the sensitivity of local *LCOA* in region 12 to the CAPEX of wind turbines and electrolyzers, and the results are shown in Fig. 8. A reduction of 30% in capacity cost of facilities in the future will lead to competitive *LCOA* for WtA.

Furthermore, according to the ammonia production in Inner Mongolia (1.06 Mt), the substitution of WtA to CtA can save approximately 1.79 Mtce of coal consumption [30] and reduce up to 4.89 Mt of $CO_2$ emission [31] per year, which is also one major advantage of WtA.

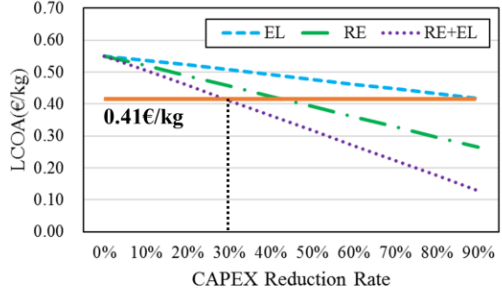

Fig. 8 Sensitivity of the local *LCOA* in region 12 to CAPEX of facilities

## V. CONCLUSIONS

This paper proposes the optimal configuration model of WtA at the network level which considers three supply modes (Local WtA, EN-based WtA and HSC-based WtA). Several significant factors are considered into the model to determine the optimal planning of WtA facilities and optimal combination of three supply modes in cases with spatial discrepancies, including the temporal fluctuation of wind resources based on the evaluation, the operation flexibility of the ammonia synthesis reactor and the transport distances. The case studies discuss the optimal planning results, optimal energy transport modes and optimal combination of three WtA modes in Inner Mongolia. Furthermore, the discussion on the effect of different wind resources on the planning capacity of WtA facilities shows that a stronger fluctuation of wind power corresponds to a larger required capacity of electrolyzers and hydrogen buffer tanks and a smaller planning capacity of wind turbines, which provides the guideline for the utilization of wind resources in different regions. Energy transport via HSC is more cost-efficient in short distances (below 500 km), but EN must be the supplement mode in long-distance energy transport. Hence, investigating the option of new pipelines for hydrogen transport over large distances in future research can create further insights. The average LCOA is 0.57 €/kg in Inner Mongolia, and the reduction of 30% in capacity cost of facilities in the future will result in competitive LCOA for WtA compared to traditional CtA (0.41 €/kg).

## APPENDIX

Fig. 9 shows the relationship of annual wind energy potential $E$ and average full-load hour (FLH) with wind turbines' capacities $P^{RE}$ in Inner Mongolia based on the evaluation results.

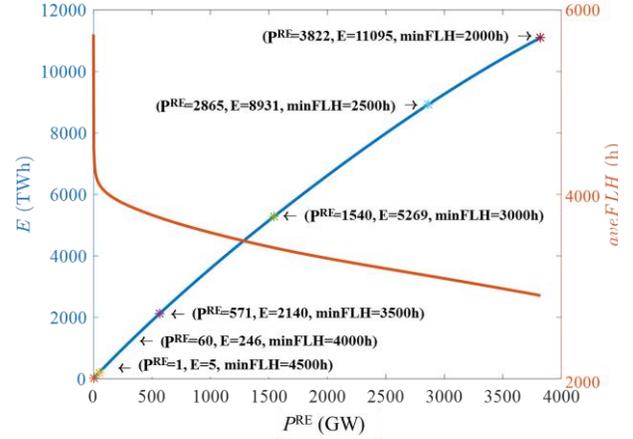

Fig. 9 Evaluation results of wind energy potential in Inner Mongolia

TABLE III
PARAMETERS IN (1) AND (2)

| region | $a_i$ (h/d/MW) | $b_i$ (h/d) | $P_i^{RE,max}$ (MW) |
|---|---|---|---|
| 2 | -1.39e-05 | 11.52 | 22585 |
| 4 | -1.58e-04 | 12.52 | 980 |
| 8 | -6.34e-05 | 11.44 | 2655 |
| 9 | -3.81e-05 | 11.50 | 2240 |
| 10 | -7.02e-04 | 12.66 | 250 |
| 11 | -3.36e-05 | 11.58 | 6820 |
| 12 | -5.49e-02 | 13.99 | 40 |

TABLE IV
ECONOMIC INDEXES OF FACILITIES [14], [25]

| Cost | Unit cost | fix*OPEX/CAPEX* | Lifetime (years) |
|---|---|---|---|
| $EX^{RE}$ | 1000 €/kW | 2% | 20 |
| $EX^{EL}$ | 500 €/kW | 3% | 10 |
| $EX^{BUF}$ | 500 €/kg $H_2$ | 2% | 20 |
| $EX^{truck}$ | 37.21 €/kg $H_2$ | 12% | 8 |
| $EX^{trailer}$ | 200 €/kg $H_2$ | 2% | 12 |

In Table V, D represents the demand regions and S represents the source regions.

TABLE V
DETAILED VALUES OF $D_{j \rightarrow i}$ (KM)

| D\S | 1 | 2 | 3 | 4 | 5 | 6 | 7 | 8 | 9 | 10 | 11 | 12 |
|---|---|---|---|---|---|---|---|---|---|---|---|---|
| 1 | 0 | 186 | 591 | 1121 | 1484 | 270 | 2006 | 437 | 238 | 1339 | 723 | 747 |
| 2 | 186 | 0 | 405 | 1301 | 1664 | 140 | 2186 | 251 | 318 | 1519 | 903 | 561 |
| 3 | 591 | 405 | 0 | 1706 | 2069 | 348 | 2591 | 154 | 723 | 1924 | 1308 | 156 |
| 4 | 1121 | 1301 | 1706 | 0 | 363 | 1391 | 1365 | 1552 | 983 | 698 | 398 | 1862 |
| 5 | 1484 | 1664 | 2069 | 363 | 0 | 1754 | 1002 | 1915 | 1346 | 335 | 761 | 2225 |
| 6 | 270 | 140 | 348 | 1391 | 1754 | 0 | 2276 | 313 | 408 | 1609 | 993 | 504 |
| 7 | 2006 | 2186 | 2591 | 1365 | 1002 | 2276 | 0 | 2437 | 1868 | 667 | 1283 | 2747 |
| 8 | 437 | 251 | 154 | 1552 | 1915 | 313 | 2437 | 0 | 569 | 1770 | 1154 | 310 |
| 9 | 238 | 318 | 723 | 983 | 1346 | 408 | 1868 | 569 | 0 | 1201 | 585 | 879 |
| 10 | 1339 | 1519 | 1924 | 698 | 335 | 1609 | 667 | 1770 | 1201 | 0 | 616 | 2080 |
| 11 | 723 | 903 | 1308 | 398 | 761 | 993 | 1283 | 1154 | 585 | 616 | 0 | 1464 |
| 12 | 747 | 561 | 156 | 1862 | 2225 | 504 | 2747 | 310 | 879 | 2080 | 1464 | 0 |